\documentclass[preprint,preprintnumbers,superscriptaddress,amsmath]{revtex4}
\usepackage{url}
\usepackage{epsfig}
\usepackage{graphicx,color}
\usepackage{epstopdf}
\usepackage[psamsfonts]{amssymb}
\usepackage{amsmath}
\usepackage{indentfirst}
\usepackage{ulem}

\begin{document}

\title{Limits of Declustering Methods for Disentangling Exogenous from Endogenous
Events in Time Series with Foreshocks, Main shocks and Aftershocks}

\author{D. Sornette}
\affiliation{Department of Management, Technology and Economics,
ETH Zurich, Kreuzplatz 5, CH-8032 Zurich, Switzerland}
\affiliation{Department of Physics, ETH Zurich}
\affiliation{Department of Earth Sciences, ETH Zurich}
\affiliation{Institute of Geophysics and Planetary Physics
and Department of Earth and Space Sciences,
University of California, Los Angeles, CA 90095}
\email{dsornette@ethz.ch}

\author{S. Utkin}
\affiliation{Mathematical Department,
Nizhny Novgorod State University, Gagarin prosp. 23,
Nizhny Novgorod, 603950, Russia}
\email{sergei_utkin@mail.ru}

\begin{abstract}
Many time series in natural and social sciences can be
seen as resulting from an interplay between exogenous influences
and an endogenous organization.
We use a simple (ETAS) model of events occurring sequentially, in which future events
are influenced (partially triggered) by past events to ask
the question of how well can one disentangle the exogenous
events from the endogenous ones.
We apply both model-dependant and model-independent stochastic declustering methods
to reconstruct the tree of ancestry and estimate key parameters.
In contrast with previously reported positive results,
we have to conclude that declustered catalogs
are rather unreliable for the synthetic catalogs that
we have investigated, which contains of the order
of thousands of events, typical of realistic applications.
The estimated rates of exogenous events
suffer from large errors. The key branching ratio $n$,
quantifying the fraction of events that have been triggered
by previous events, is also badly estimated in general from declustered catalogs.
We find however that the errors tend to be smaller
and perhaps acceptable in some cases
for small triggering efficiency and branching ratios. The high level of randomness together with
the very long memory makes the stochastic
reconstruction of trees of ancestry and the estimation
of the key parameters perhaps intrinsically unreliable for
long memory processes. For shorter
memories (larger ``bare'' Omori exponent), the results improve
significantly.

\end{abstract}

\date{\today}

\maketitle

\section{Introduction}

A large variety of natural and social systems are characterized by a stochastic intermittent
flow of sudden events: landslides, earthquakes, storms, floods, volcanic eruptions, biological extinctions, traffic
gridlocks, power blackouts, breaking news, commercial blockbusters,
financial crashes, economic crises, terrorist acts, geopolitical events, and so on.
Sequences of such sudden events constitute often the most crucial
features of the evolutionary dynamics of complex systems, both in terms
of their description, characterization and understanding. Accordingly,
a useful class of models of complex systems views their dynamics as
a sequence of intermittent discrete short-lived events. In the limit
where the time scales, over which the change of regimes associated with the
occurrence of the events occur, are small compared with the inter-event
intervals, the catalog of events can be modeled using the mathematics
of point-processes \cite{Bremaud1,Daley_VJ}. This modeling strategy emphasizes that the system is active during
short-lived events and inactive otherwise. This amounts to separating
a more or less incoherent background activity (such as small
undetectable earthquakes) from the occurrence of
structured events (large earthquakes), which are the focus of interest.
We note that the class of stochastic point processes is fundamentally different
from that of discrete and continuous stochastic processes, for which the
activity is non-zero most of the time.

Having a time series or catalog of discrete events, we are interested in understanding
the generating process that led to the observed sequence. The difficulty in
deciphering the underlying mechanisms
stems from the fact that the above systems of interest are on the one hand subjected
to external forcing which on the other hand provide them the stimuli to self-organize via negative and positive
feedback mechanisms.
Most natural and social systems are indeed continuously subjected to external stimulations, noises, shocks, solicitations,
and forcing, which can widely vary in amplitude. It is thus not clear a priori if the observed activity is due to a strong
exogenous shock, to the internal dynamics of the system organizing in response to the continuous flow of information and
perturbations, or maybe to a combination of both. In general, a combination of external inputs and internal organization
is at work and it seems hopeless to disentangle the different contributions to the observed collective human response.
Determining the chain of causality for such questions requires disentangling interwoven exogenous and endogenous
contributions with either no clear or too many signatures. How can one assert with confidence that a given event or
characteristics is really due to an endogenous self-organization of the system, rather than to the response to an external
shock?

It turns out that a significant understanding of the complex flow of observed events can
be achieved by precisely framing the problem in
terms of a classification of two limited classes of events:
(i) those that are the response of the system to exogenous
shocks to the system and (ii) those that appear endogenously without any
obvious external causes. This can be done by
looking as the specific endogenous and exogenous signatures and their
mutual relations, which are reminiscent of the fluctuation-susceptibility theorem
in statistical physics \cite{ruellefluc,sornette2005origins}. This approach provides a useful framework for understanding
many complex systems and has been successfully applied in several contexts:
commercial book successes \cite{sornette2004amazon,deschatres2005amazon}, social crises \cite{Roehneretal04}, financial
volatility \cite{MRWvol}, financial bubbles and crashes \cite{wtmc_Princeton03,johsortestendoexo}, earthquakes
\cite{HSG03,HStrigger03}, diseases in
complex biological organisms \cite{sor-endo-illness09}, epileptic seizures \cite{Osorioetal} and so on.

A common feature observed in these different systems is the fact that events are
not independent as they would be if generated by a Poisson process. Instead,
they exhibit pronounced inter-dependencies, characterized by ``self-excitation'', i.e.,
past events are found to often promote or trigger (in part) future events,
leading to epidemic-like cascades of events. The analogy with triggering
and cascade processes occurring in viral epidemics is so vivid
in some instances that the name ``Epidemic-type
aftershock model'' (ETAS) has been given to one of the most popular model
of earthquake aftershock processes \cite{Ogata1988,HSbasic02,HSfore}. The ETAS model
belongs to the class of self-excited conditional point processes introduced
in mathematics by Hawkes \cite{Hawkes1,Hawkes2,Hawkes3}. It constitutes
an excellent first-order approximation to describe the spatio-temporal organization of
earthquakes \cite{Saichevsor07} and is now taken as a standard
benchmark. This class of ``self-excited'' point processes
also provides quantitative predictions on the different decay rates after exogenous peaks
of activity on the one hand
and endogenous peaks of activity on the other hand. These
predictions have been verified in a unique data set of
almost 5,000,000 time series of human activity collected sub-daily
over 18 months since April 2007 from the 3$^{rd}$ most visited web site YouTube.com.

Given these preliminary successes, we would like to decipher, understand
and perhaps forecast the dynamics of events. For this, it is important to recognize that
the observed dynamics can be modeled as a complex entangled mixture of events,
from exogenous shocks impacting the systems providing
bring surprises, that are progressively endogenized by the system, which
is also capable of purely endogenous (or internal) bursts of activity.
In between, real systems can be viewed as organized by a mixture of
exogenous shocks and endogenous bursts of activity. The grail is to
disentangle these different types of events.

The purpose of the present paper is to contribute a step towards
the operational problem of
disentangling the exogenous and endogenous contributions to the organization
of a system revealed by a time series of discrete events. For this, we use
synthetic catalogs of events generated by the ETAS model. This model
describes the occurrence of successive events, each of them characterized
by its time $t_i$ of occurrence and a ``mark'' $M_i$, that we refer to as ``magnitude'' to borrow
from the vocabulary of earthquakes. Generally, the mark can be any trait or property that
influence the ability of the event to trigger other future events.
The conditional
intensity $\lambda(t,M|H_t)$ of the linear version of the general self-excited Hawkes process (which we use
later to define the specifics of the ETAS model) reads
  \begin{equation}\label{lambda}
    \lambda(t,M|H_t)=\mu(t,M)+\sum\limits_{t_i<t}
    h(t,M;t_i,M_i)~.
  \end{equation}
$\lambda(t,M|H_t)$ consists of two main contributions:
(i) the background intensity $\mu$ which, if alone, would give
a pure Poisson process of background events (more complicated
source terms can be considered, but we keep $\mu$ constant
in the present paper); (ii) the response
functions $h$, one for each past event, describing the propensity to trigger future events.
Specifically, $h(t,M;t_i,M_i)$ is the intensity
of the $i$-th event that occurred at time $t_i$ with magnitude $M_i$ to produce an aftershock at
a later time $t>t_i$ with magnitude $M$. $H_t$ stands for the history known at time
$t$, and thus includes the sequence of all events from the beginning of observations till $t$.

Zhuang et al.  \cite{Zhuangetal02,Zhuangthesis} have proposed a rigorous stochastic
declustering method that, in essence, implements the program of disentangling the
different events according to their background (exogenous) or triggered (endogenous)
origins.  Zhuang et al. applied the space-time ETAS model to real earthquake catalogs.
The main problem is that no validation test was performed to check if the
reconstruction of the ancestry sequence was realistic and if the inverted parameters
were consistent, i.e., without bias. In the present paper, our goal is to make
a systematic investigation of the quality of the stochastic reconstruction method(s).
For this, using the ETAS model (\ref{lambda}), we generate known synthetic catalogs of events, which
are considered as our ``observations''.  We then apply
different approaches, which are variants of the ``stochastic declustering'' methods
introduced by Zhuang et al.  \cite{Zhuangetal02,ZhuangOgata04}
and generalized by Marsan and Lenglin\'e \cite{Marsan07}. Our strategy
is to compare the key parameters obtained from
the reconstruction of the cascade of triggering events obtained by these
declustering methods applied to our ``observations'' to the known true values. This allows us to  establish the
uncertainties and biases associated with these ``model inversions''.
As we shall see, the level of stochasticity, inherent in self-excited
conditional Poisson processes, introduces rather dramatic errors, which
appear to have been largely under-estimated in the literature. Because of the
general application of this problem to a variety of domains, we focus
our attention to time series of events, assuming that spatial information
is either irrelevant or not available. This makes the problem of stochastic
declustering less constrained than in the case of earthquakes, for instance,
in which one does have some information on the spatial positions, in addition
to the times of occurrence.

The paper is organized as follows. The next section \ref{gtwtrog2}
reviews the two methods of stochastic declustering. Section \ref{thtaqaqer}
first describes the ETAS model that we have used for the generation of
synthetic catalogs. Then, it presents the two specific implementations
of the ETAS model, referred to respectively as the generation-by-generation
and event-by-event algorithms. After recalling the main results of previous
tests performed by other authors, Section \ref{htogjjqq} defines the
parameters that are tested and presents the main results using
the two declustering SDM and MISD methods on synthetic catalogs
generated by the two generation-by-generation and event-by-event algorithms.
Section \ref{thjtyohjpqfeq} concludes.

\section{Description of different declustering variants \label{gtwtrog2}}

The general idea underlying a stochastic declustering method applied to a sequence
of events thought to be generated by a self-excited conditional Poisson process
is to attribute to each event a probability of being either an exogenous (a so-called ``background'' event)
or an offspring of previous events (endogenous). Obviously the former probability is one minus
the later probability. Therefore the main task is to estimate one of these probabilities for each event.
Specifically, it is convenient to focus on the probability that a given event is a background event.

We start from the stochastic declustering method introduced
in Ref.~\cite{Zhuangetal02,ZhuangOgata04}. This method can be implemented
under several technical variants,
such as the thinning (random deletion) method, and a variable bandwidth
kernel function method associated with the maximum likelihood estimation
of the ETAS model. Using any of these two approaches, the background intensity is obtained
by using an iterative algorithm, and the declustered catalogs can
also be generated.

\subsection{Zhuang et al's declustering method \cite{Zhuangetal02,ZhuangOgata04} \label{sec_thin}}

We focus on the thinning procedure, which uses
the probabilities $\rho_{ik}$ for the $k$-th event
to be an aftershock of the $i$-th event and the probability $\phi_k$ that
the $k$-th event is only a background event. These probabilities can be expressed
in terms of the response function and the intensity defining the ETAS model (\ref{lambda}):
  \begin{equation}\label{rho_ik}
    \rho_{ik}=\frac{h(t_k,M_k;t_i,M_i)}
    {\lambda(t_k,M_k|H_t)}~.
  \end{equation}
This allows us to define the probability $\rho_{k}$
that the $k$-th event is an aftershock  (whatever its triggering ``mother'') as
\begin{equation}\label{rho}
    \rho_{k}=\sum\rho_{ik}~.
  \end{equation}
Therefore, the probability $\phi_k$ that
the $k$-th event is only a background event is
 \begin{equation}\label{phi}
    \phi_k=1-\rho_{k}=\frac{\mu(t_k,M_k|H_{t_k})}
    {\lambda(t_k,M_k|H_t)},
  \end{equation}
 These probabilities are  called ``thinning probabilities''  \cite{Zhuangthesis}.

  If we delete the $k$-th event in the catalog with probability $\rho_k$
  for all $k = 1; 2; \ldots ;N$, then the thinned process should realize
  a non-homogeneous Poisson process of the intensity $\mu(t,M)$
  (see \cite{Ogata81} for the mathematical justification). This process is called the
  background sub-process, and the complementary sub-process is the cluster
  or offspring process. The following algorithm implements this thinning procedure.

\subsection*{Algorithm 1}

The indices $k = 1; 2; \ldots ;N$ of the events in the catalog are
ordered according to their time sequence: $t_i < t_2< ...< t_N$.
  \begin{enumerate}
  \item For all events $k = 1, 2, \ldots,N;$, calculate their
probabilities $\rho_k$ in (\ref{rho});

  \item Generate N uniform random numbers $U_1, U_2, \ldots, U_N$ in
$[0; 1]$.

  \item If $U_k < 1 - \rho_k$, keep the $k$-th event; otherwise, delete
it from the catalog as an offspring. The remaining events can be
regarded as the background events.
\end{enumerate}
This algorithm can be applied to any data series and will find
thinning probabilities for each event, which are functions of the specific model used.

The next issue is to estimate the parameters defining the response functions
$h(t_k,M_k;t_i,M_i)$ of the Hawkes point process. For this, the
following algorithm determine
which event in the data set is the ancestor of a given $k$-th event.

\subsection*{Algorithm 2}

\begin{enumerate}
\item For each pair of events $i; k = 1, 2, \cdots , N (i <
k)$, calculate the probability $\rho_{ik}$ in (\ref{rho_ik}) and
$\phi_k$ in (\ref{phi}).

\item Set k = 1.

\item Generate a uniform random number $U_k \in [0; 1]$.

\item If $U_k < \phi_k$ , then the $k$-th event is regarded as
a background event.

\item Otherwise, select the smallest index $I$ in $\{i+1,...,N\}$ such that $U_k < \phi_k +
\sum\limits_{i=1}^I \rho_{ik}$. Then, the $k$-th event is regarded
to be a descendant of the $I$-th event.

\item If $k = N$, terminate the algorithm; else set $k = k + 1$
and go to Step 3.
\end{enumerate}
The output of this algorithm is to provide a complete classification
of events as backgrounds or descendent of some previous event,
this previous event being either a background or a descendent
of another previous event and so on. The corresponding reconstructed
ancestry tree allows us to calculate different parameters of the model,
such at the productivity law, giving the average number of offsprings
of a given event as a function of its magnitude. Of course, a given
ancestry tree obtained by this stochastic declustering method is not unique since,
given a fixed catalog of events,  it depends on the realization of the random
numbers $\{U_k\}$ in Algorithms 1 and 2. Thus, these stochastic reconstructions
must be performed many times with different independent realizations
of the random numbers $\{U_k\}$ to obtain many statistically equivalent
ancestry trees over which statistical averages can be taken.

  \subsection{Marsan and Lenglin\'e's Model-Independent Stochastic Declustering (MISD) \cite{Marsan07}}

The MISD method also aims at determining the thinning probabilities
(\ref{rho_ik}) \cite{Marsan07,misd_url}, using a rapidly converging
algorithm with a minimum set of hypotheses. The MISD method proposes to
reveal the full branching structure of the triggering process, while avoiding
model-dependent inversions.

The MISD method has the following key assumptions:
\begin{itemize}
  \item The generating process of the observed catalog of events is considered to be a point process, in time, space
      and magnitude (we will only consider the situation where no spatial information is provided in this paper to
      focus only on time series). Its conditional intensity consists of a linear superposition of a constant (Poisson
      process) background intensity $\mu$ and of the branching part represented by the summation in the
r.h.s. of the following expression,
      \begin{equation}
        \lambda(t,\bar x, m)=\mu+\sum\limits_{t_i<t} \lambda_i(t_i,\bar x_i, m_i)~,
      \end{equation}
      where the sum is performed over all $i$-th events that have occurred before time $t$.

  \item The average activity of events in response to the occurrence of a given event only depends on its magnitude
      $m$.
\end{itemize}
While very general, it is however necessary to point out that the MISD method assumes that the conditional Poisson
intensity  $\lambda_i(t_i, \bar x_i, m_i)$ due to each passed event $i$ contributes additively (linearly)
to the total conditional intensity $ \lambda(t,\bar x)$. This assumption is appropriate if the generating
process belongs to the class of linear self-excited Hawkes processes (\ref{lambda}).
However, this linearity condition excludes a large class of nonlinear self-excited conditional Poisson models
\cite{Bremaud2,Bremaud3}, and in particular the class of multiplicative (in contrast to additive)
Poisson models \cite{Multifractal1,Multifractal2}  endowed with general multifractal scaling properties
\cite{SaichevSormultifract}.

\textbf{The MISD algorithm} includes two iterating steps (that we present in full generality, including
the possible existence of a spatial information):
\begin{enumerate}
  \item Starting with a first a priori guess of the ``bare'' kernel $\lambda(t,\bar x, m)$ and of $\mu$, the
      triggering weight (thinning probability) is estimated by $\rho_{ij}=a_j\lambda(t_j-t_i,|\bar x_j - \bar x_i|,
      m_i)$ for $t_i<t_j$ and 0 otherwise. The ``background weight'' is estimated as $\phi_j=a_j\mu$, where $a_j$ is a
      normalization
coefficient.

  \item The updated (a posteriori) ``bare'' rates are then computed as
  \[
  \lambda(|\Delta t|,|\Delta \bar x|,m) = \frac{1}{N_m \times \delta t \times S(|\Delta \bar x|. \delta r)} \sum_{i,j
  \in A} \rho_{ij},
  \]
where $A$ is the set of pairs such that
$|\bar x_i-\bar x_j|=|\Delta\bar x|\pm \delta r$, $m_i=m\pm \delta m$ and $t_j-t_i=t\pm \delta t$ ($\delta r$, $\delta
t$, $\delta m$ are
discretization parameters), $N_m$ is the number of events such that $m_i=m\pm \delta m$,
and $S(|\Delta \bar x| \delta r)$ is the surface covered by the disk with radii $|\Delta\bar x|\pm \delta r$.
Similarly, the a posteriori background rate is
\[
\mu = \frac{1}{T \times S} \sum\limits_{j=1}^{N} \phi_j,
\]
where $T$ is the duration of the time
series (containing $N$ events) and $S$ is the surface analyzed.
\end{enumerate}

Roughly speaking, the first step of the algorithm « selects » the
triggering events for each triggered event (i.e., it
assigns triggering weights based on our present knowledge of the
rates). The second step then updates these rates, using the
intermediate branching structure obtained at the first step. The solution
is accepted (convergence is achieved) when
the a priori and the a posteriori kernels
are identical, implying that the rates and weights are
consistent with each other.


The initial formulation recalled above and the implementation of the
MISD algorithm was done for three- and four-dimensional data,
including time, magnitude and spatial coordinates of events
\cite{Marsan07,misd_url}.  Lately, D. Marsan has adapted the code
to two-dimensional data (times and magnitudes) and we use his code
for the research reported here. In our implementation, following
D. Marsan and others, we use logarithmic binned time and linear magnitude bins.


\section{Model and simulation implementations \label{thtaqaqer}}

This section describes the specific ETAS model that we use to generate
synthetic time series of events, that are then collated in catalogs
on which the different stochastic declustering algorithms described
in the previous section are applied. Previous implementation by
Zhuang et al.  \cite{Zhuangetal02,ZhuangOgata04} have used the ETAS
model with its full space-time formulation. Here, we use the ETAS
model in which the spatial information is supposed to be non-existent
or irrelevant. In this way, our tests are relevant to catalogs of events
obtained in other systems, such as social, commercial  financial, and biological systems.

\subsection{The ETAS Model}
\label{sec_model}

The ETAS model is defined as the self-excited linear conditional
Poisson model with intensity (\ref{lambda}), in which the (``bare'') response function
$h(t,M;t_i,M_i)$ is expressed at the product of three terms
\begin{equation}\label{os_ik}
h(t,M,t_i,M_i|H_{t_k}) = j(M) \cdot \Phi(t_i|H_{t_k}) \cdot Q(M_i)~,
\end{equation}
where
\begin{equation}\label{gr_}
  j(M) = b~ \ln 10\cdot 10^{-b(M-m_0)}~,
\end{equation}
has the form of a Gutenberg-Richter law prescribing the frequency
of events of magnitudes $M$,
\begin{equation}\label{omori}
  \Phi(\tau) = \frac{\theta c^\theta}{(c+\tau)^{1+\theta}}
\end{equation}
has the form of the Omori-Utsu law specifying the
PDF (probability density function) of time intervals between a main event and its direct
aftershocks, and
\begin{equation}\label{prod}
  Q(M) = K ~ 10^{\alpha M}
\end{equation}
is the productivity law giving the mean number of direct aftershocks
generated by an event as a function of its magnitude $M$.
We thus have
\begin{equation}\label{os_ik3}
h(t,M,t_i,M_i|H_{t_k}) = b~ \ln 10\cdot 10^{-b(M-m_0)} \cdot
\frac{\theta c^\theta}{(c+t-t_i)^{1+\theta}} \cdot K 10^{\alpha M_i}~.
\end{equation}
Equations (\ref{os_ik},\ref{os_ik3}) express the
independence between the determination of the magnitude
of aftershocks and their occurrence times on the one hand, and with the
magnitudes of their triggering ancestors on the other hand.
Next, the background rate $\mu(t,M)$ in (\ref{lambda}) is taken also
multiplicative as
\begin{equation}
\label{nhyhb2y}
\mu(t,M)= j(M) \cdot  \mu~.
\end{equation}
The constant $\mu$ means that the background events are occurring according to a standard
memoryless Poisson process with constant intensity.
The multiplicative structure of $\mu(t,M)$ in (\ref{nhyhb2y}) again
expresses that the magnitudes of the background events are
independent of their occurrence times.

In summary, the conditional intensity of the ETAS model used here reads
\begin{equation}\label{ourlambda}
\lambda(t,M|H_{t_k}) = j(M) \cdot \left(\mu +
\sum\limits_{i=1}^{k} \Phi(t_i|H_{t_k}) \cdot Q(M_i)\right)~,
\end{equation}
with the definitions (\ref{gr_},\ref{omori},\ref{prod}) and the constant
background rate $\mu$.
Substituting (\ref{nhyhb2y}) and  (\ref{ourlambda})
into (\ref{rho}) and (\ref{phi}), we get the sought
thinning probabilities for the ETAS model.

In order to generate synthetic catalogs of events, we considered
two different simulation algorithms. Both techniques
used the following set of parameters:
\begin{equation}\label{para}
  \Theta = (c, \theta, \alpha, b, n)
\end{equation}
where
\begin{equation}
\label{nwhbbnta}
n = \frac{K}{1-\alpha/b}
\end{equation}
is the branching ratio -- the mean
number of direct aftershocks per triggering event \cite{HSbasic02}.
The first algorithm keeps the information about
the mother-daughter relations by generating events generation by generation. It
uses a slightly different implementation of the ETAS model than
the second algorithm and is  computationally more costly in the
sense that a generated catalog becomes stationary only after tens or even hundreds of thousands events.
The second algorithm, which is based on the formulation of the
model described in \ref{sec_model}, is not very fast but is efficient.
It generates events that belong to the stationary regime from the beginning, so no information about preceding events is
lost (unlike the first algorithm). However, it loses the information about the mother-daughter relations. Its performance
was validated in Ref.~\cite{SorSaiUtk08}.

\subsection{First algorithm (generation-by-generation): Generation of synthetic catalogs keeping the information on the
relations between events}\label{algo_f}

The first simulation algorithm we use here was developed by K. Felzer
\cite{Felzer02}. The idea is to generate events generation-by-generation.
Firstly, the mother-shock and the background events are generated,
then goes the first generation of aftershocks of existing events,
after that the second generation and so on. The procedure stops when
the time boundary or the limit on the number of events is reached. We
slightly modified the algorithm (mostly input and output) to
correspond to our needs.

The advantage of knowing the ancestry relations
between events, which allows us to  estimate
parameters such as $\alpha$ and $b$, comes at the cost
of the existence of a long transient before the time
series of events become stationary. Exactly at criticality $n=1$, where
$n$ is given by (\ref{nwhbbnta}), this transient becomes infinitely
long lived since the renormalized Omori law \cite{HSbasic02}, which takes into account
all generations of events triggered by each event, develops
an infinite memory in the technical sense \cite{Beran} with a
non-integrable decay $1/t^{1-\theta}$, where $\theta$ is defined in (\ref{omori}). This
makes this second algorithm unreliable for $n$ close to $1$.

\subsection{Second algorithm (event-by-event): Generation of synthetic catalogs without any information on the relations
between events}\label{algo_m}

Compared with standard
numerical codes that have been used
by previous workers to generate synthetic catalogs of
events, the event-by-event algorithm uses the specific formulation of the
ETAS model and of its known conditional cumulative distribution
function (CDF) of inter-event times. Thus, parts of the
calculations can be done analytically. The occurrence
times are generated one-by-one by determining the CDF of the time till the next event
based on the knowledge of the previous CDF and the time of the
previous event (firstly introduced by Ozaki for Hawkes' processes \cite{Ozaki79}). For this, we use a standard
Newton algorithm as well as a
standard randomization algorithm. The event-by-event algorithm
is not very fast because it has to solve the equation for the CDF numerically each time. However,
as already mentioned above, its advantage stems from the fact that the generated events
belong to the stationary regime from the beginning of each catalog.

Let us recall how the CDFs $F_{k+1}(\tau)$ for $k=2, 3, ...$ of successive events are obtained recursively.
The first event is supposed to have occurred at time $t_1=0$.

The CDF $F_2(\tau)$ of the waiting time from the first to the
second shock is made of two contributions: (i) the second shock
may be a background event or (ii) it may be triggered by the
first shock. This yields
\begin{equation}
F_2(\tau)=1-e^{-\omega\tau} e^{-q_1(1-a(\tau))}~,
\label{eq_second}
\end{equation}
where $q_1=Q(m_1)$ is the productivity of the first shock obtained
from expression (\ref{prod}) given its magnitude $m_1$, and
$a(\tau)$ is defined as
\begin{equation}\label{at}
a(\tau)= \int_\tau^\infty \Phi(t') dt' = {c^\theta \over (\tau +
c)^\theta}~ .
\end{equation}

All following shocks are similarly either a background event or
triggered by one of the preceding events. The CDF $F_3(\tau)$ of
the waiting time between the second and the third shocks is thus
given by
\begin{equation}
F_3(\tau)=1-e^{-\omega\tau} e^{-q_1(a(t_2-t_1)-a(t_2+\tau-t_1))
-q_2(1-a(\tau))}~,
\end{equation}
where $t_2$ is the realized occurrence time of the second shock. Iterating, we obtain the CDF $F_k(\tau)$ for
the waiting time between the $(k-1)$-th and $k$-th shocks under the
following form
\begin{equation}
F_k(\tau)=1-e^{-\omega\tau} \exp\left[-\sum_{i=1}^{k-1}
q_{i}\cdot \left(a\left(t_{k-1}-t_i\right)
-a\left(t_{k-1}+\tau-t_i\right)\right)\right]~,
\label{eq_k}
\end{equation}
where $t_j$ is the occurrence time of the $j$-th event (which is equal to sum of all generated time intervals between
events prior to the $j$-th one), and $q_i=Q(m_i)$ is the productivity of the $i$-th shock obtained from expression
(\ref{prod}) given its magnitude $m_i$.

In order to generate the $(k+1)$-th inter-event time interval
between the occurrence of the $k$-th and $(k+1)$-th shock, it is
necessary to know the $k$ previous inter-events times between the
$k$ previous shocks and their $k$ magnitudes. Since, in the ETAS
model, the magnitudes are drawn independently  according to the
Gutenberg-Richter distribution (\ref{gr_}), they can be generated
once for all. In order to generate a catalog of $N$ events, we
thus draw $N$ magnitudes from the law (\ref{gr_}). In order to
generate the corresponding $N$ inter-event times, we use
expression (\ref{eq_k}) iteratively from $k=1$ to $k=N$ in a
standard way: since any CDF $F(x)$ of a random variable $x$  is by
construction itself uniformly distributed in $[0,1]$, we obtain a
given realization $x*$ of the random variable $x$ by drawing a
random number $r$ uniformly in $[0,1]$ and by solving the equation
$F(x*)=r$. In our case, we generate $N$ independent uniformly
distributed random numbers $x_1, ..., x_N$ in $[0,1]$ and
determine each $\tau_i$ successively as the solution of
$F_i(\tau_i)=x_i$.

As mentioned above, the main shortcoming of this procedure is that
it does not record if an event was spontaneous or
a descendant of some previous event. For that reason, we cannot
use it for all parameter estimations.

\subsection{Preliminary tests of the synthetic catalogs}

We have checked the consistency of our algorithms by verifying that, for $K=0$
in (\ref{prod}) corresponding to the absence of triggering,
a Poisson flow of event time occurrence is obtained.

\begin{figure}[htp]
\includegraphics[width=12cm]{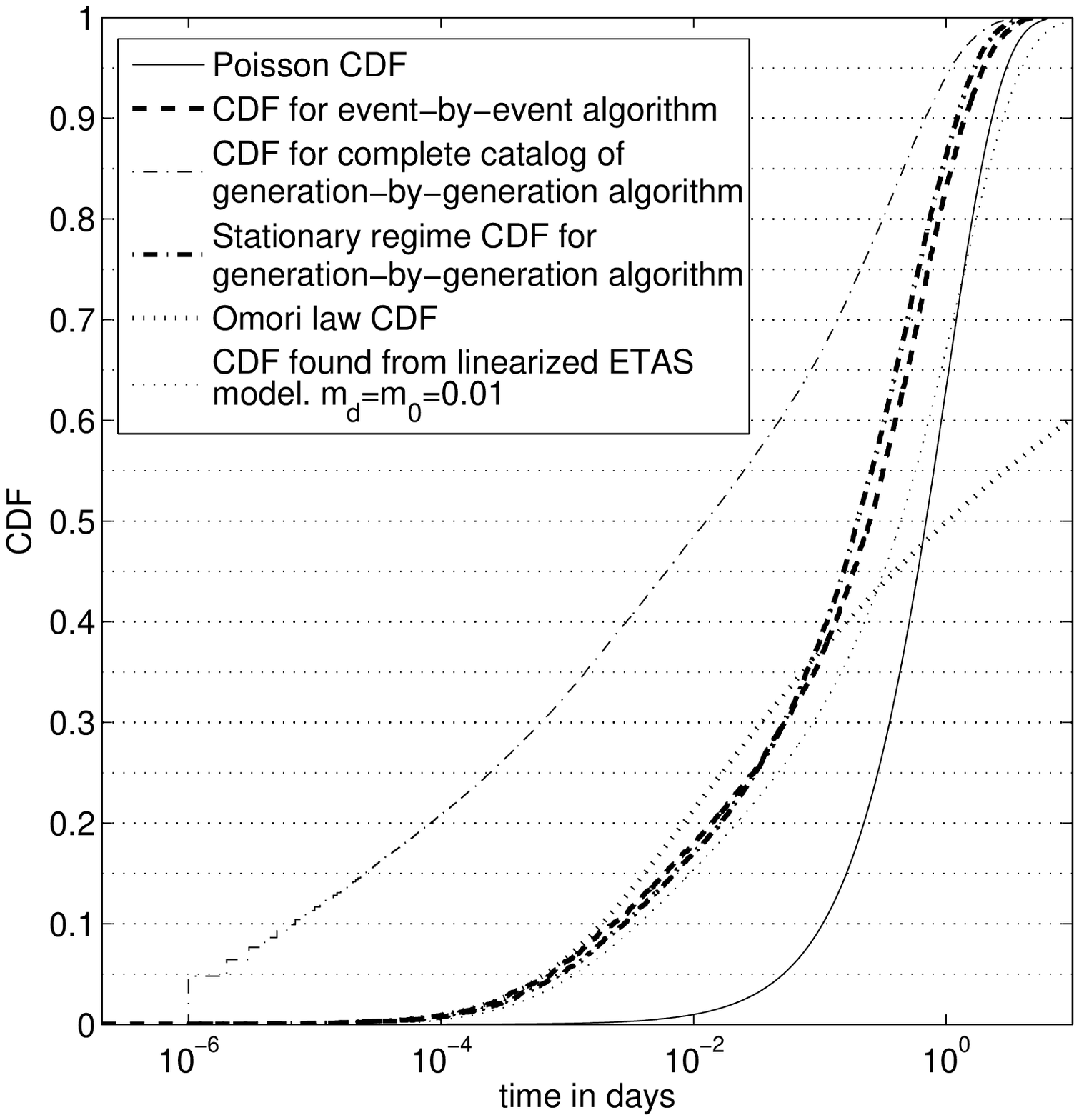}
\caption{CDFs of inter-event times for synthetic catalogs generated
for $b=1$, $c=10^{-3}$, $\alpha=0.7$, $\theta=0.1$, $n=0.7$ ($K=0.21$) and $m_d-m_0=0.01$.
The CDFs of inter-event times are shown for the two algorithms (first event-by-event \ref{algo_m}
and second generation-by-generation \ref{algo_f}) and for
the Poisson flow of the background events, the bare Omori law (\ref{omori})
and  the theoretical prediction obtained from the linearized equation of the ETAS model
developed in \cite{SaichevSor06,Saichevsor07}.}
\label{test_sim}
\end{figure}

We have implemented the two algorithms just discussed in the previous subsection
and have constructed the corresponding CDFs from the obtained time series
for  $b=1$, $c=10^{-3}$, $\alpha=0.7$, $\theta=0.1$, $n=0.7$ ($K=0.21$) and $m_d-m_0=0.01$.
Figure \ref{test_sim} shows that both algorithms lead to CDFs which are very
close to each other, when the transient regimes of the catalogs generated
by the first generation-by-generation algorithm are removed. Using the whole catalog including
the transient part for the first algorithm leads
to very large distortions. We interpret the remaining slight difference between
the CDFs of inter-event times for the first and second
algorithms after removal of the transient as due to the residual influence of the transient part
of the catalog in the first generation-by-generation algorithm  \ref{algo_f}.
Figure \ref{test_sim} also shows for comparison the inter-event CDFs of
(i) the Poisson flow of the background events, (ii) the bare Omori law (\ref{omori})
and (iii) the theoretical prediction obtained from the linearized equation of the ETAS model
developed in \cite{SaichevSor06,Saichevsor07} confirming the need to go to the
nonlinear description  \cite{SorSaiUtk08}.

\section{Results and performance of stochastic declustering methods (SDM and MISD) \label{htogjjqq}}

\subsection{Previous tests of Zhuang et al.  \cite{ZhuangOgata04,Zhuangetal02,OgaZhu06} }

As mentioned before, Zhuang et al. \cite{ZhuangOgata04,Zhuangetal02,OgaZhu06} applied their
declustering procedure described in section \ref{sec_thin}
to real earthquake catalogs over four geographical regions: New Zealand (NZ),
Central and Western Japan (CJ and WJ) and Northern China (NC).
Table \ref{tab_zh} provides the results of their SDM applied to these four regions.

Our main remark is that no error or uncertainty analysis is reported and no study of
the impact of the lower magnitude threshold used in the catalog is performed.
This is particularly worrisome, given the demonstration that parameter estimations
are significantly biased when the minimal observable (registerable) magnitude $m_d$
is different from (usually larger than) the minimal event magnitude $m_0$ able to produce aftershocks
 \cite{Sorwerner,SaiSor06}.

 In a later paper, Zhuang et al.
\cite{Zhuangetal08} reported some synthetic tests to assess the reliability of the SDM
in the ideal case where $m_d=m_0$. They
quoted ``good reconstruction results''. However, the parameters estimated with their SDM
($c\approx 0.0004,\; \alpha\approx 0.57,\; \theta\approx 0.014,\; n\approx 0.25$)
were very different from the true parameters
($c=0.0002,\; \alpha\approx 0.65,\; \theta=0.12,\; n\approx 0.99$) used to generate
the synthetic catalogs. Very worrisome is the very large error in the value
of the branching ratio $n$. The true value $n=0.99$ corresponds to a system
close to critical branching in which triggering of multiple generations is expected
to be very strong since $99\%$ of events are triggered on average while only $1\%$
are exogenous \cite{HStrigger03}.
In contrast, the estimated value $n \approx 0.25$ would be interpreted
as a relatively weak triggering regime in which three-quarters of the events
are exogenous.

\subsection{Previous tests of Marsan and  Lenglin\'e \cite{Marsan07}}

Unlike the SDM of Zhuang et al, the MISD method was partially verified.
Marsan and  Lenglin\'e  generated synthetic catalogs with the
parameters $b=1$, $c=0.01$, $\alpha=0.87$, $\theta=0.2$, $n=0.9$ and $\mu=0.25$.
Applying the MISD to that catalog, Marsan and  Lenglin\'e
could estimate the background rate  $\mu^{est}=0.248\pm 0.01$, very close to the
true value.  The estimates of the other parameters were reported to be also good \cite{Marsan07} .

\subsection{Self-consistency of parameter estimations of $n$ and $\alpha$ from synthetic catalogs}

\subsubsection{A modifications to SDM (henceforth referred to as mSDM)}

In the above presentation, we have considered only the thinning probabilities of events in a catalog.
But Zhuang et al's method contains in addition a determination of
the conditional intensity function (\ref{ourlambda}) obtained by
using an iterative search procedure with the maximum likelihood $L(\Theta)$ \cite{Daley_VJ}.
Specifically, the search procedure determines the set of
parameters $\Theta$ of the model for which the thinning probabilities are best consistent with the conditional intensity
$\lambda(t,M|H_{t_k})$.  Zhuang et al. used the first order algorithm of Davidson-Fletcher-Powell
to find the sought maximum of the likelihood function. This method suffers from the need
to calculate the derivatives of $\lambda(t,M|H_{t_k})$ and $L(\Theta)$, which makes errors accumulate
and slows down the calculations.

In the present work, we use the Nelder-Mead simplex method which is significantly more
efficient than the first-order Davidson-Fletcher-Powell algorithm. In addition, we recalculate
the probabilities $\rho_{ij}$ for each estimation of the likelihood function rather than at each
iteration as performed in the SDM version used by Zhuang et al. Our approach increases
the computation time of a single iteration but actually provides a significant gain as
the number of iterations needed for convergence is greatly reduced.

The set of parameters $\Theta^d$ obtained from the maximization of the likelihood function corresponds to the
\textit{direct estimation}. The priority is to verify whether the direct estimation is good enough. Then, we need to
check
that the catalog reconstructed using thinning probabilities also provides good estimates of the model parameters.

\subsubsection{Applying the modified SDM (mSDM) to two-dimensional data}

 To check the goodness of the direct estimation by the mSDM, we generated 5 catalogs of 2500 events for different values
 of $\alpha$ and $n$ and fixed $c=0.001$, $b=1$ and $\theta=0.5$. As one can see from the results
 presented in table \ref{tab_add3}, the direct estimation can be quite far off, in particular for the
 parameters $\theta$ and $n$. But the errors turn out to be smaller than with the thinning probabilities
 that will be presented below. One of the origins for the errors is the limited length
of the synthetic catalog, nothwithstanding the use of an intensionally large value of $\theta=1/2$
leading to a rather short memory (compared to that for smaller values of $\theta$ used below).
The SDM and mSDM need more events to reduce the errors in the parameter estimation.
Further below, we will consider the relationship between the length of a catalog that is
needed to obtain reasonable results and the memory
quantified by the exponent $\theta$).

\subsubsection{Targeted parameters}

As a first test, we assume known the parameters of the ETAS process generating
the synthetic catalogs. For instance, we can assume that the
use of the mSDM led us to the true values of parameters
$\Theta^d \approx \Theta^s$. We then apply the SDM and the MISD to determine
the background and triggered events. From this knowledge, we can estimate
directly the branching ratio $n$ and the fertility exponent $\alpha$ and compare
them with the true values.  We stress that we use the exact parameters that enter in the generation of the synthetic
catalogs to find the thinning probabilities (\ref{rho_ik}) to perform this test.
Thus, any discrepancy between the estimation of $n$ and $\alpha$
using the thinning probabilities should be
attributed only to errors in the reconstruction of the tree of ancestry through the thinning probabilities.

The first key parameter of interest is the branching ratio, which can be estimated
from the knowledge of the number of exogenous events within the data set, according to
\begin{equation}\label{n}
n_e = 1 - \frac{{\rm number\; of\; background\; events}}{\rm total\; number\; of\; events}~,
\end{equation}
where the
subscript \textit{e} indicates that $n_e$ is experimental or estimated value.

Knowing the tree structure of events obtained from the declustering method,
we can estimate directly the productivity law (\ref{prod}). Specifically,
the tree structure allows us to calculate straightforwardly the mean number of direct aftershocks
triggered by a given event, and then to test how this mean number depends on the
magnitude of the mother event. Given the true law (\ref{prod}), the estimated
dependence of the number of aftershocks as a function of the magnitude of the main shock
is fitted by the following expression
\begin{equation}\label{prod_f}
  \hat Q(M) = K^*\cdot 10^{A^*\cdot M}~,
\end{equation}
where $(K^*,A^*)$ are determined using
standard optimization algorithms. The estimated values can then be compared to
the true values $(K,\alpha)$ used to generate the synthetic catalogs.

To account for the fact that, in real time series, small events below
a magnitude detection threshold $m_d > m_0$ are not detected, we also
investigate the influence of this detection threshold on the estimated
parameters. Intuitively, as demonstrated in Refs.~\cite{Sorwerner,SaiSor06},
missing events lead to misinterpret triggered events as exogenous
background events, since the chain of causal triggering may be ruptured.
This may influence severely the estimation of the background rate
and therefore of the parameters controlling the fertility and triggering efficiency
of past events. A good diagnostic of the effect of missed events
is the branching ratio $n$ \cite{Sorwerner,SaiSor06}. We thus evaluate
the apparent branching ratio estimated by the declustering method
on the catalog of events with magnitudes larger than $m_d$ according
to the following formula
\begin{equation}
  n_t = \frac{\rm number\; of\; aftershocks\; with\; M>m_d,\; whose\; mother\; also\; has\; M>m_d}{\rm number\; of\; all\;
  events\; with\; M > m_d}
  \label{hjrwptbjr2pt}
\end{equation}

We also will test how the truncation affects the estimated parameters, and does this correspond to the theoretical
dependance \cite{SaichevSor06,Saichevsor07}.

\subsubsection{First test on declustering  Poisson sequences}

We first applied the SDM and MISD to catalogs generated by
a simple Poisson process, obtained from the formulation of section
\ref{sec_model} by imposing the value $K=0$ in (\ref{prod}).
As a consequence, only exogenous background events without
any triggered event are generated in synthetic catalogs with
intensity imposed equal to $\mu=1$. A correct declustering
algorithm should find $n=0$ and a background rate ${\hat \mu} = 1$.

Applying the SDM on tens of catalogs each containing
between 1000 and 2000 events with magnitudes $M>0$,
we recovered the correct result that the branching ratio is $n=0$ in every case
(all probabilities $\phi_j$ are found exactly equal to 1) and the
background rate is estimated as ${\hat \mu} = 1\pm 0.05$, close and
consistent with the true value.

On a technical note, the arrest criterion controlling the convergence
of the algorithm was found to play a strong role, much more
significant for time-domain-only catalogs than when the catalogs
include spatial information. We found these
good results only for
an arrest criterion corresponding to differences smaller than $0.0001$
between the parameter values of successive iterations.  We kept this
value for all subsequent tests, as a compromise between 
accuracy of the convergence and numerical feasibility.

On the same catalogs, the MISD method found however nonzero
probabilities $\rho_{ij}$ for $i \neq j$ and, as a result, a nonzero
and actually quite large
branching ratio \textbf{$n = 0.23 \pm 0.03$}. The estimation of the background rate
was also not very good ${\hat \mu}_0 = 0.81 \pm 0.02$ instead of the
true value $1$. This means that the MISD method applied in the
temporal domain to already declustered catalogs (pure Poisson)
misclassifies about \textbf{$23\%$}  of the events as being triggered,
while they are all exogenous.

\subsubsection{Tests using catalogs generated by the
generation-by-generation algorithm \ref{algo_f}}\label{sec_res1}

{\bf Test of the SDM}. We tested the SDM using data sets generated with various
values of $\alpha= 0\; -\; 0.9$ of the productivity law (\ref{prod}).
We generated 10 catalogs of length of $\approx 50,000$ days
for each parameters set, with a background rate $\mu_0=1$ per day.
The duration of the catalogs in terms of days is just to offer a convenient
interpretation, as the intrinsic time scale is
more generally determined by $1/\mu_0$.
Each catalog contained about $70,000\; -\;
100,000$ events. We removed the first $2000$ events in the
first part of the catalogs, roughly corresponding to the first
1500 days, and applied the SDM procedure 20 times to each
catalog. Table \ref{tab_NewFAlpha} compiles the obtained values
$n_e$, $A^*$ and $K^*$ defined in (\ref{n}) and (\ref{prod_f}),
reports their standard deviations and compares with the
true values $n, \alpha$ and $K$.

We found that the distributions of background rates
for the different values of  $\alpha= 0\; -\; 0.9$ are reasonably
estimated, with an error of no more than $10\%$.
While the estimated branching ratio $n_e$ is also reasonably close to the
true value for $\alpha$ up to $0.6$ and then
starts to systematically deviate for larger $\alpha$'s,  the estimated parameters
$K^*$ and $A^*$ are found very far from the true values. In particular,
the values of the estimated fertility exponent $A^*$ would imply
that events of all magnitudes have on average no more than one aftershock,
which is very far from being the case, especially for large values of $\alpha$.
One partial cause for this bad result is the Omori law which,
as shown in fig.\ref{test_sim},  implies very long
inter-event times between direct aftershocks. Some of those intervals can
be longer than our catalog and the real productivity will be thus
underestimated. Another possible partial cause is that the removal of
an initial part of the catalogs to analyze the more
stationary regime at later times deletes by construction many events that are
mothers of the observed events. This also leads to an
underestimation of the triggering productivity. These two
explanations suggest that the problem is intrinsic
to the application of the SDM to the ETAS model and we
do not envision easy fixes.

{\bf Test of the MISD method}. We applied the MISD method
to the same catalogs. Table  \ref{tab_misd19}
shows a slight systematic overestimation of the branching ratio $n_e$
over the true value $n$. The estimations of $\alpha$ and $K$
are also bad with large systematic errors. The trend
of variation of $K$ as a function of $\alpha$ for the fixed true $n=0.26$
is qualitatively reproduced by the dependence of $K^*$ as
a function of $\alpha$.

We must also report a surprising difference between
the SDM and the MISD method. While the standard deviations
of the estimated parameters over 10
different synthetic catalogs are sometimes significantly larger for the MISD
method compared with the SDM, the former method exhibited
sometimes very accurate results for a few catalogs for some
specific values of the parameters. For instance, for one of the
synthetic catalog generated with the
parameters $n=0.26$, $\alpha=0.9$, $K=0.026$ and $\mu_0=1$, the MISD method
gave the following estimates: $n_e=0.255 \pm 0.007$, $A^*=0.916 \pm
0.037$, $K^*=0.018 \pm 0.007$ and background rate ${\hat \mu}_0 = 0.890
\pm 0.218$. The existence of such an excellent inversion has to be tampered
by the fact that the estimates obtained with the MISD method applied
to the other 9 catalogs generated
with the same parameters were bad. This suggests a very strong
dependence of the performance of the MISD method on the specific stochastic realizations.

{\bf Impact of catalog incompleteness}. We now report some results
on the estimation of parameters by the two declustering methods applied
to incomplete catalogs, motivated by the nature of real-life catalogs.
The incompleteness is measured by the
magnitude threshold $m_d > m_0$, below which events
are missing from the catalogs used for the SDM and MISD method.

\begin{figure}[htp]
\includegraphics[width=12cm]{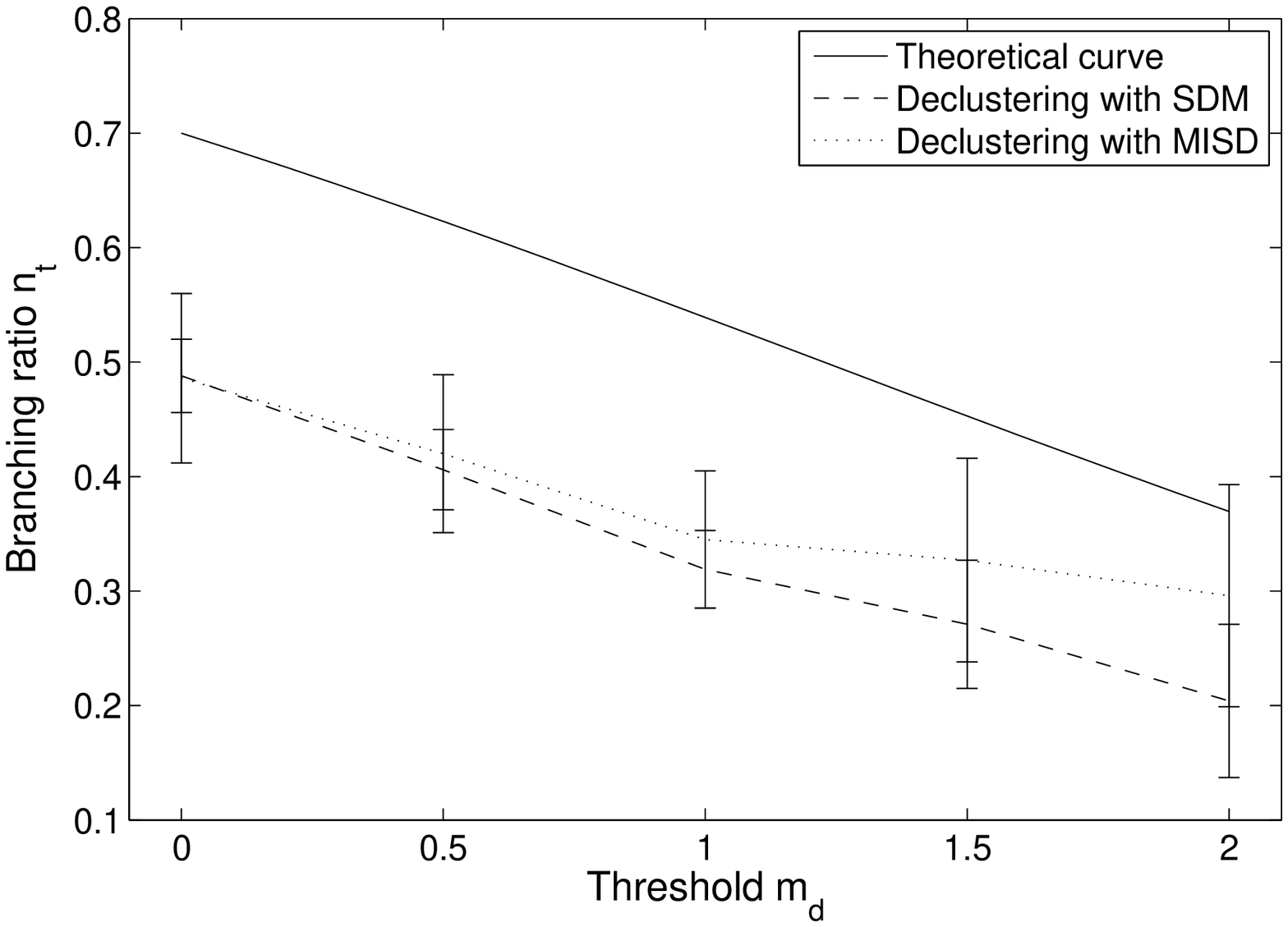}
\caption{Dependance of the effective branching ratio $n_t$
as a function of the threshold magnitude $m_d$ of catalog incompleteness.
The parameters used to generate the synthetic catalogs with the ETAS model
are $b=1,\; c=0.001,\; \theta=0.1,\; n=0.7,\; \alpha=0.7,\; K=0.21$.
The dashed (respectively dotted) line correspond to $n_t$
obtained by using SDM algorithm (respectively MISD algorithm).
The continuous curve is the validated theoretical formula (\ref{eq_nt_md}). }
\label{fig_nmd20}
\end{figure}

First, we focus on the results  \cite{Sorwerner,SaiSor06}
that the branching ratio $n$ is renormalized into an effective value $n_t$
which is a decreasing function of $m_d$:
\begin{equation}
\label{eq_nt_md}
  n_t(m_d) = \frac{1}{1+\frac{1-n}{n} [10^{\alpha m_d}]^{{b \over \alpha}-1}}~.
\end{equation}
This prediction was verified by direct simulations with the ETAS model. Here,
we test how the incompleteness of catalogs may interfere with
the stochastic declustering methods.
We generated 50 catalogs with the following set of parameters:
$b=1,\; c=0.001,\; \theta=0.1,\; n=0.7,\; \alpha=0.7,\; K=0.21$
and varied $m_d$ from $0$ to $2$. Applying the SDM
20 times to each of the 50 synthetic
data sets, figure \ref{fig_nmd20} shows that (i) the general trend of decreasing
$n_t(m_d)$ as a function of $m_d$ is recovered,
but (ii) there is a very significant downward bias of approximately $0.2$
over the whole range $0 \leq m_d \leq 2$.
Table \ref{tab_NewFMd} shows unsurprisingly that the estimated
parameters $A^*$ and $K^*$ are very far from the true values $\alpha$
and $K$. Using incomplete catalogs cannot be expected to
improve the estimation of parameters which is already bad
for complete catalogs. For smaller values of the true branching ratio $n$,
the discrepancy is smaller between the reconstructed $n_t$
and the theoretical formula (\ref{eq_nt_md}). For instance, for $n=0.4$,
the difference between the estimated $n_t$ and the formula (\ref{eq_nt_md})
decreases from $0.1$ for $m_d=0$ (no incompleteness) to
almost zero for $m_d=2$ for which $n_t(m_d=2) \approx 0.04$.

Typical results for the MISD method are reported in
table \ref{tab_misd20} for the set of true parameters $n=0.7, \alpha=0.7$
and $K=0.21$, for $m_d$ varying from $0$ to $2$. While
the estimated $n_e$ for $m_d=0$ (no incompleteness) are as good as with
the direct SDM estimation method,
the other estimated parameters are strongly biased. For incomplete
catalogs $m_d>0$, we observe a large over-estimation of $n_t$
and significant errors in the other estimated parameters.

\begin{figure}[htp]
\includegraphics[width=12cm]{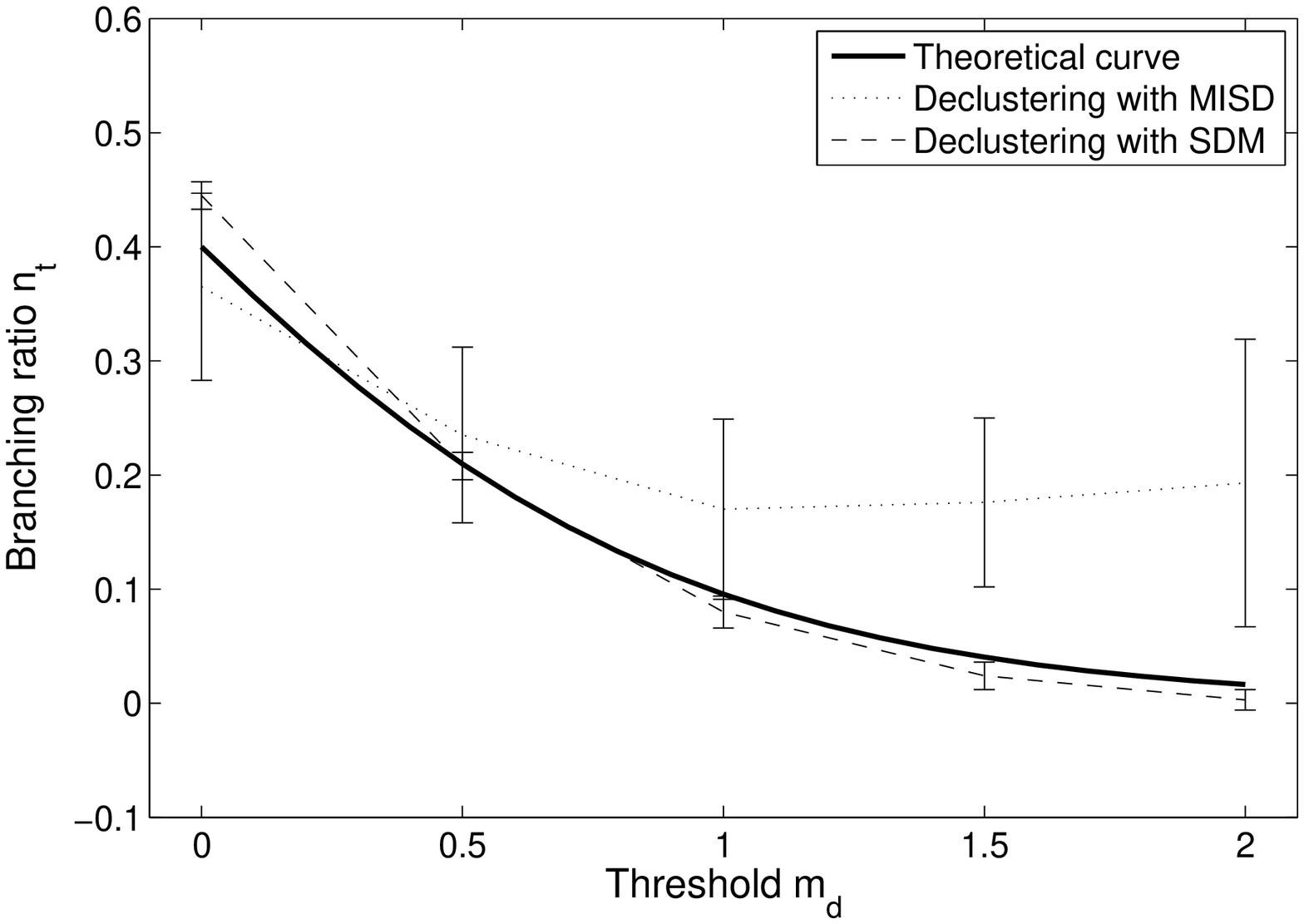}
\caption{Dependance of the effective branching ratio $n_t$
as a function of the threshold magnitude $m_d$ of catalog incompleteness.
The parameters used to generate the synthetic catalogs with the ETAS model
are $b=1,\; c=0.001,\; \theta=0.1,\; n=0.4,\; \alpha=0.2,\; K=0.32$.
The dashed (respectively dotted) line correspond to $n_t$
obtained by using the SDM algorithm (respectively the MISD algorithm).
The continuous curve is the validated theoretical formula (\ref{eq_nt_md}). }
\label{fig_nmd21}
\end{figure}

Another test is provided by comparing the CDFs of background events
in the incomplete catalogs as a function of $m_d$ obtained
by the declustering methods with the true CDFs. We found that,
the larger is the true branching ratio $n$, the larger is the
discrepancy between the true and reconstructed CDFs of
background events, using both declustering methods
for all $m_d$ values. For $n \leq 0.4$, the reconstructed
background CDFs are in reasonable agreement
with the theoretical formula (\ref{eq_nt_md}) with typical errors
of about $10\%$ (See tables \ref{tab_misd21} and \ref{tab_NewFMd} and figure \ref{fig_nmd21}).  For large branching
ratios, the errors are too large and the declustering methods are unreliable.

Comparing figures \ref{fig_nmd20} and \ref{fig_nmd21}, one can notice that MISD becomes less precise for large thresholds
$m_d$. That is more likely caused by the smaller lengths of the catalogs.

\subsubsection{Tests using catalogs generated by the event-by-event algorithm \ref{algo_m}}

The same tests as reported in the previous subsection were performed
on catalogs generated by the event-by-event algorithm \ref{algo_m}.
We recall that our motivation for using this alternative algorithm
is to test for the expected influence of transient regimes, which are
absent by construction in the synthetic catalogs obtained with
the event-by-event algorithm \ref{algo_m}.

Using one hundred catalogs of 2000 events (10 for each
$\alpha$ going from $0$ to $0.9$) generated with algorithm \ref{algo_m},
we applied the SDM and obtained the results summarized in table
\ref{tab_NewMAlpha}. The results are similar to those of table \ref{tab_NewFAlpha},
with a reasonable estimation $n_e$ but estimated $A^*$ and $K^*$
very far from the true values $\alpha$ and $K$.

Ten catalogs of 4000 events were generated with the parameters
$b=1, c=0.001, \theta=0.1, n=0.6, \alpha=0.2$ and $K=0.48$.
Incompleteness was introduced at magnitude thresholds $m_d$
varying from $0$ to $2$, reducing the size of catalogs to the
$m_d$-dependent number  $\approx N_{m_d}$ events. Applying
the SDM to these incomplete catalogs, we obtained the results
shown in table \ref{tab_NewMMd}. As mentioned above,
the estimated $n_e$ is found in better agreement with the
theoretical prediction (\ref{eq_nt_md}) for the largest $m_d$ values
for which $n_t$ is the smallest.

Table \ref{tab_misd21} shows that the MISD method gives
results similar to those previously obtained in table \ref{tab_misd20}.
The estimated value of background rate is
${\hat \mu}_0 = 0.165 \pm 0.198$, which is quite different
from the true value $1$.


\section{Tests on the influence of memory (exponent $\theta$), background rate and catalog lengths}

\subsection{Influence of the value of the memory exponent $\theta$}

We now test one possible origin for the rather bad
performance of the SDM and MISD method when using the thinning probabilities, namely the
very long memory quantified by the small value of the
exponent $\theta$ (as defined in expression (\ref{omori})) used in the simulations.

A series of tests were made with a larger value of $\theta=0.5$, corresponding
to significantly shorter memory.  We varied the values of $n$ and $\alpha$,
while fixing the other parameters $c=0.001$, $b=1$, $\mu=1$.
We generated 10 catalogs of 2500 events each using the event-by-event algorithm
described in subsection \ref{algo_m} for each set of parameters.
We implemented both declustering algorithms 20 times to each catalog.
Table \ref{tab_add1} presents the resulting estimates of the parameters.
The most striking result is that the branching ratio $n$ estimated
by the SDM is in general very good. While the MISD method is not
reliable for estimated the branching ratio $n$, it is better for the
estimation of the productivity exponent $\alpha$, especially
for large values.  The results presented in table \ref{tab_add1},
when put in comparison with those of the previous tables obtained with
much smaller values of the memory exponent $\theta$,
illustrates clearly the impact of the long-memory on the declustering results.

\subsection{Case of a single large main shock triggering aftershocks}

In another series of tests, we check if the SDM or MISD are able to determine a pure tree branching process
emanating from a single source. In this goal, we removed all spontaneous events except one, the first ``main shock''. This
corresponds to imposing $\mu=0$. In order to have sufficiently many events in the catalogs, we take
the magnitude of the single main shock large ($M_1=7$) and also impose large values for
the parameters $n$ and $\alpha$ to produce long sequences of events.
The parameter $\theta$ was taken equal to 0.1 (long memory).
We found that the SDM recognized the existence of the only background event in two tests out of three
(for $\alpha=0.8$). In contrast, the MISD lacks efficiency in such conditions and
proposes significantly non-zero values for the background rate $\mu$.
The estimation of $n$ with formula (\ref{n}) cannot provide good results because it should give $n_e \to 1$ (only one
background event) for large catalogs. 
Other methods of estimating the
branching ratio described in \cite{HStrigger03} gave even worse results, with $n_e>1$.

\subsection{Influence of catalog lengths}

\begin{figure}[htp]
\includegraphics[width=12cm]{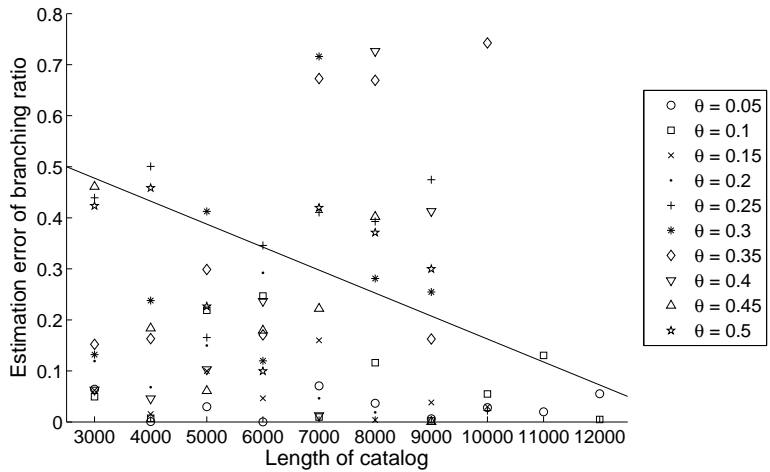}
\caption{Error of the direct estimation of the branching ratio by the mSDM as a function of catalog length for different
values of $\theta$.}
\label{fig_ern}
\end{figure}

As mentioned above, two effects combine to limit the efficiency of the SDM and MISD methods:
the smallness of the exponent $\theta$ leading to very long memory and the limited length
of the catalogs. We investigated how these two effects are inter-related in practice.
Using algorithm \ref{algo_f}, we generated 100 catalogs with variable numbers of events (from $3000$ to $12000$) and for
various values of $\theta$ (from $0.05$ up to $0.5$). The other parameters were fixed at $b=1$, $\alpha=0.7$, $c=0.001$,
$n=0.7$. We then applied the mSDM to each catalog.
Fig.\ref{fig_ern} shows the estimation error of the branching ratio $n$. One can observe
very large variations from realization to realization and a weak tendency for estimation errors
to decrease with the length of the catalogs.

\begin{figure}[htp]
\includegraphics[width=12cm]{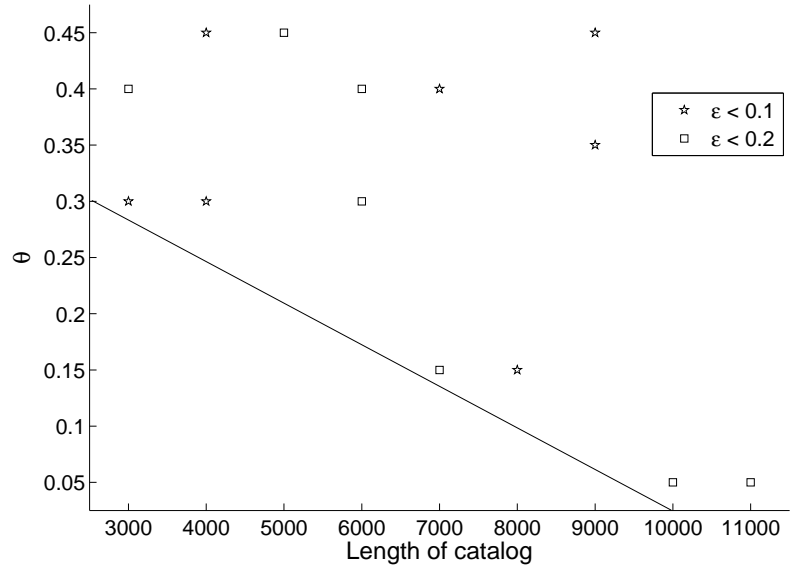}
\caption{Dependence of $\varepsilon$ as a function of catalog lengths is shown for
two classes of catalogs sorted according to their estimation errors as defined
by equation (\ref{eq_cumerr}).}
\label{fig_ntt}
\end{figure}

To be more quantitative, let us introduce the cumulative error ratio defined as the sum of squares of relative errors
over all parameters:
\begin{equation}\label{eq_cumerr}
  \varepsilon = \sum\limits_{i=1}^{5} \left( 1 - \frac{\Theta_i^d}{\Theta_i^s} \right)^2,
  \quad (\Theta_{1-5}=b, \alpha, c, \theta, n).
\end{equation}
The dependence of $\varepsilon$ as a function of catalog lengths is shown in
fig.\ref{fig_ntt} for a number of realizations. Again, one can observe an overall decrease
of the estimation error with the length of the catalogs, decorated by a very large variability
from catalog to catalog.

As an illustration of the quality of the reconstruction of the tree structure
of ancestry in different catalogs, we took 7 catalogs with $\varepsilon<0.1$ (best results) and several catalogs with
$\varepsilon>1$ (bad results) and determined the percentage of background events recognized as background and of
aftershocks recognized as aftershocks. For catalogs with the best directly estimated parameters, the percentage were 66\%
and 72\% respectively, while for the ``bad'' catalogs the results were 9\% and 93\% (i.e., almost all events were
incorrectly recognized as aftershocks).

\section{Conclusions  \label{thjtyohjpqfeq}}

Many time series in natural and social sciences can be
seen as embodying an interplay between exogenous influences
and an endogenous organization. We have used a simple
model of events occurring sequentially, in which future events
are influenced (partially triggered) by past events to ask
the question of how well can one disentangle the exogenous
events from the endogenous ones.

The exogenous events
are modeled here by a Poisson flow of so-called
background events with constant intensity $\mu_0$. The ETAS
specification of the
conditional self-excited Hawked Poisson model has been used.
It contains three principal ingredients: (i) a long Omori-like
power law memory of the influence of past events on future events,
(ii) a Gutenberg-Richter-like distribution of event magnitudes and (iii) a
fertility law expressing how many events are triggered by a given
event as a function of its magnitude.

In order to separate
background events from triggered events, we have implemented
and compared two so-called ``declustering'' algorithms, the SDM introduced
by Zhuang et al. \cite{Zhuangetal02,Zhuangthesis}
and the MISD method proposed by
Marsan and Lenglin\'e \cite{Marsan07}.
We have applied these two methods to synthetic catalogs
generated by two algorithms using the ETAS model,
the generation-by-generation algorithm \ref{algo_f}
and the event-by-event algorithm \ref{algo_m}.

We specifically address the problem of reconstructing
the tree of ancestry of the sequence of events in 
recorded catalogs. In particular, one main goal is to distinguish
the exogenous shocks (background events) from the endogenous
(triggered events).  For these problems, we find  
that declustered catalogs obtained from synthetic catalogs generated with the ETAS model
are rather unreliable, when using catalogs with of the order
of thousands of events, typical of realistic applications.
The estimated rates of exogenous events
suffer from large errors. The key branching ratio $n$,
quantifying the fraction of events that have been triggered
by previous events, is also badly estimated in general with these approaches.
We find however that the errors tend to be smaller
and perhaps acceptable in some cases
for the smaller fertility exponent $\alpha <0.6$ and for
the smaller branching ratio $n<0.4$ typically. Results
become better when the memory exponent $\theta$ is increased,
i.e., when the memory is shortened.
We do not find significantly better performance
of the SDM versus the MISD method or vice-versa, with the curious
observation that the MISD method is sometimes
very precise for some catalog realizations, but this
property is not robust with respect to other stochastic
realizations. We have also investigated the role
of incompleteness on declustering, and found that
this is not the essential limiting problem.

We should however make clear that these rather
negative results are not necessarily opposed to the
more positive results reported by Zhuang et al.  and
Marsan and Lenglin\'e, which refer to a different problem,
that of the direct estimation of the model parameters (and not
of the tree of ancestry and of the distinction between exogenous
and endogenous events).
Our larger ambition to reconstruct the tree of ancestry 
has identified clearly intrinsic limits of the inversion process.

It appears
that the high level of randomness together with
the very long memory makes the stochastic
reconstruction of trees of ancestry and the estimation
of the key parameters quite unreliable. Technically,
we find the coexistence of many coexisting stochastic
reconstructions with different parameter estimates, and it is not obvious how to select what
should be the right one. This question is reminiscent
of complex optimization problem in the presence
of a very large number of almost equivalent solutions,
as occurs in so-called NP-complete problems. There thus
appears to be fundamental limitations intrinsic
to this class of models.

\vskip 1cm
{\bf Acknowledgements}:  We are grateful to David Marsan for discussions
and help in using his MISD algorithm as well as feedbacks on the
testing procedure. We thank Anne-Marie Christophersen, David Marsan
Guy Ouillon, Max Werner and Jiancang Zhuang for useful
feedbacks on an earlier version of the manuscript.

\newpage

\clearpage

\begin{table}[htp]
\caption{Estimated parameters of the ETAS model obtained by
the stochastic declustering method (SDM) of
Zhuang et al. \cite{ZhuangOgata04,Zhuangetal02,OgaZhu06}  described in section \ref{sec_thin}
applied to real earthquake data from New Zealand (NZ),
Central and Western Japan (CJ and WJ) and Northern China (NC). The
quoted values for the fertility exponent $\alpha$ differ from those reported
by Zhuang et al. by the conversion factor $\ln 10$ accounting for our
use of base-ten logarithm and exponential compared with the natural logarithm
and exponential used by Zhuang et al. }
\label{tab_zh}
  \begin{tabular}{ccccccc}
  Region &\vline& $c$  & $\theta$ & $\alpha$ & $K$   & $n$   \\
  \hline
  NZ &\vline& 0.017    & 0.164    & 0.389    & 0.335 & 0.548 \\
  CJ, WJ &\vline& 0.040& 0.250    & 0.495    & 0.204 & 0.404 \\
  NC &\vline& 0.003    & 0.030    & 0.499    & 0.546 & 1.090\\
  \hline
\end{tabular}
\end{table}

\begin{table}[htp]
\caption{Direct estimation of the parameters $\Theta^s$ with the modified SDM compared with the true parameters
$\Theta^d$ of the ETAS model used to generate the synthetic catalogs by the event-by-event algorithm.}
\label{tab_add3}
  \begin{tabular}{rcccccccc}
  \# &            &\vline& $b$ & $\alpha$ &   $c$   & $\theta$ & $n$ \\
  \hline
  1  & $\Theta^s$ &\vline& 1.00 &    0.20 & 0.00100 & 0.50    & 0.20 \\
     & $\Theta^d$ &\vline& 3.80 &    0.27 & 0.00103 & 0.54    & 0.07 \\
  \hline
  2  & $\Theta^s$ &\vline& 1.00 &    0.50 & 0.00100 & 0.50    & 0.50 \\
     & $\Theta^d$ &\vline& 0.44 &    0.30 & 0.00050 & 0.21    & 0.78 \\
  \hline
  3  & $\Theta^s$ &\vline& 1.00 &    0.80 & 0.00100 & 0.50    & 0.80 \\
     & $\Theta^d$ &\vline& 1.00 &    0.96 & 0.00160 & 0.64    & 0.96 \\
  \hline
  4  & $\Theta^s$ &\vline& 1.00 &    0.20 & 0.00100 & 0.50    & 0.80 \\
     & $\Theta^d$ &\vline& 0.33 &    0.24 & 0.00240 & 0.84    & 0.92 \\
  \hline
  5  & $\Theta^s$ &\vline& 1.00 &    0.80 & 0.00100 & 0.50    & 0.20 \\
     & $\Theta^d$ &\vline& 1.00 &    0.76 & 0.00070 & 0.41    & 0.10 \\
  \hline
\end{tabular}
\end{table}

\begin{table}[htp]
\caption{Estimated parameters $n_e$, $A^*$ and $K^*$ with the SDM
compared with the true parameters $n=0.26$, $\alpha$ and $K$ of the ETAS model
used to generate the synthetic catalogs by the generation-by-generation algorithm. The other parameters have been
fixed to $m_d=m_0=0$, $b=1$,
$c=0.002$, and $\theta=0.19$.}
\label{tab_NewFAlpha}
  \begin{tabular}{ccrccccc}
  \# &\vline&  $n$  &   $n_{e}$   & $\alpha$ & $A^*$  &  $K$  & $K^*$ \\
  \hline
  1  &\vline&  0.26 & $0.237 \pm 0.009$ & 0.0 & $0.024 \pm 0.041$ & 0.260 & $0.930 \pm 0.097$ \\
  2  &\vline&  0.26 & $0.239 \pm 0.010$ & 0.1 & $0.015 \pm 0.060$ & 0.234 & $0.939 \pm 0.118$ \\
  3  &\vline&  0.26 & $0.240 \pm 0.010$ & 0.2 & $-0.005 \pm 0.041$ & 0.208 & $0.960 \pm 0.087$ \\
  4  &\vline&  0.26 & $0.237 \pm 0.009$ & 0.3 & $-0.048 \pm 0.039$ & 0.182 & $1.034 \pm 0.079$ \\
  5  &\vline&  0.26 & $0.240 \pm 0.012$ & 0.4 & $-0.059 \pm 0.061$ & 0.156 & $1.030 \pm 0.134$ \\
  6  &\vline&  0.26 & $0.240 \pm 0.018$ & 0.5 & $-0.092 \pm 0.053$ & 0.130 & $1.065 \pm 0.115$ \\
  7  &\vline&  0.26 & $0.226 \pm 0.012$ & 0.6 & $-0.104 \pm 0.073$ & 0.104 & $1.082 \pm 0.142$ \\
  8  &\vline&  0.26 & $0.225 \pm 0.022$ & 0.7 & $-0.084 \pm 0.150$ & 0.078 & $1.043 \pm 0.208$ \\
  9  &\vline&  0.26 & $0.207 \pm 0.029$ & 0.8 & $0.004 \pm 0.453$ & 0.052 & $0.989 \pm 0.370$ \\
  10 &\vline&  0.26 & $0.143 \pm 0.028$ & 0.9 & $-0.151 \pm 0.043$ & 0.026 & $1.097 \pm 0.068$ \\
  \hline
\end{tabular}
\end{table}

\begin{table}[htp]
\caption{Estimated parameters $n_e$, $A^*$ and $K^*$ with the MISD
compared with the true parameters $n=0.26$, $\alpha$ and $K$ of the ETAS model
used to generate the synthetic catalogs by the generation-by-generation algorithm.
The other parameters have been
fixed to $m_d=m_0=0$, $b=1$, $c=0.002$, and $\theta=0.19$.}
\label{tab_misd19}
  \begin{tabular}{ccrccccc}
  \# &\vline&  $n$  &   $n_{e}$   & $\alpha$ & $A^*$  &  $K$  & $K^*$ \\
  \hline
  1  &\vline&  0.26 & $0.319 \pm 0.048$ & 0.0 & $0.332 \pm 0.265$ & 0.260 & $0.931 \pm 0.450$ \\
  2  &\vline&  0.26 & $0.299 \pm 0.035$ & 0.1 & $0.465 \pm 0.535$ & 0.234 & $1.077 \pm 1.106$ \\
  3  &\vline&  0.26 & $0.328 \pm 0.064$ & 0.2 & $0.352 \pm 0.264$ & 0.208 & $0.925 \pm 0.693$ \\
  4  &\vline&  0.26 & $0.343 \pm 0.061$ & 0.3 & $0.488 \pm 0.373$ & 0.182 & $0.847 \pm 1.047$ \\
  5  &\vline&  0.26 & $0.336 \pm 0.062$ & 0.4 & $0.355 \pm 0.134$ & 0.156 & $0.955 \pm 0.678$ \\
  6  &\vline&  0.26 & $0.332 \pm 0.069$ & 0.5 & $0.579 \pm 0.265$ & 0.130 & $0.373 \pm 0.372$ \\
  7  &\vline&  0.26 & $0.322 \pm 0.060$ & 0.6 & $1.011 \pm 0.912$ & 0.104 & $0.293 \pm 0.325$ \\
  8  &\vline&  0.26 & $0.320 \pm 0.113$ & 0.7 & $0.815 \pm 0.488$ & 0.078 & $0.295 \pm 0.312$ \\
  9  &\vline&  0.26 & $0.337 \pm 0.151$ & 0.8 & $0.818 \pm 0.223$ & 0.052 & $0.122 \pm 0.128$ \\
  10 &\vline&  0.26 & $0.320 \pm 0.174$ & 0.9 & $0.900 \pm 0.449$ & 0.026 & $0.126 \pm 0.143$ \\
  \hline
\end{tabular}
\end{table}

\begin{table}[htp]
\caption{Estimated parameters $n_e$, $A^*$ and $K^*$ with the SDM
compared with the true parameters $n=0.7$, $\alpha=0.7$ and $K=0.21$ of the ETAS model
used to generate the synthetic catalogs by the generation-by-generation algorithm,
for various magnitude threshold $m_d$
of incompleteness. The other parameters have been fixed to
$b=1$, $c=0.001$, and $\theta=0.1$. $N_{m_d}$ is the number of events
in the incomplete catalogs.}
\label{tab_NewFMd}
  \begin{tabular}{cccrccc}
  \# &\vline&  $m_d$  & $N_{m_d}$ & $n_{e}$    & $A^*$    & $K^*$ \\
  \hline
  1  &\vline&  0.0 & 2000 & $0.488 \pm 0.032$  & $-0.045 \pm 0.400$ &  $1.037 \pm 0.214$ \\
  2  &\vline&  0.5 & 2000 & $0.406 \pm 0.035$  & $0.135 \pm 0.784$ &  $1.085 \pm 0.437$ \\
  3  &\vline&  1.0 & 1000 & $0.319 \pm 0.034$  & $-0.006 \pm 0.286$ &  $1.201 \pm 0.491$ \\
  4  &\vline&  1.5 &  500 & $0.271 \pm 0.056$  & $-0.054 \pm 0.116$ &  $1.376 \pm 0.574$ \\
  5  &\vline&  2.0 &  150 & $0.204 \pm 0.067$  & $-0.047 \pm 0.094$ &  $1.495 \pm 0.769$ \\
  \hline
\end{tabular}
\end{table}

\begin{table}[htp]
\caption{Estimated parameters $n_e$, $A^*$ and $K^*$ with the MISD method
compared with the true parameters $n=0.7$, $\alpha=0.7$ and $K=0.21$ of the ETAS model
used to generate the synthetic catalogs by the generation-by-generation algorithm,
for various magnitude threshold $m_d$
of incompleteness. The other parameters have been fixed to
$b=1$, $c=0.001$, and $\theta=0.1$. $N_{m_d}$ is the number of events
in the incomplete catalogs.}
\label{tab_misd20}
  \begin{tabular}{cccrccc}
  \# &\vline&  $m_d$  & $N_{m_d}$ & $n_{e}$    & $A^*$    & $K^*$ \\
  \hline
  1  &\vline&  0.0 & 4000 & $0.486 \pm 0.074$  & $0.930 \pm 0.448$ &  $0.167 \pm 0.166$ \\
  2  &\vline&  0.5 & 1300 & $0.420 \pm 0.069$  & $0.873 \pm 0.878$ &  $0.089 \pm 0.188$ \\
  3  &\vline&  1.0 &  400 & $0.345 \pm 0.060$  & $-0.357 \pm 2.804$ &  $0.239 \pm 0.497$ \\
  4  &\vline&  1.5 &  130 & $0.327 \pm 0.089$  & $0.249 \pm 1.521$ &  $0.168 \pm 0.686$ \\
  5  &\vline&  2.0 &   40 & $0.296 \pm 0.097$  & $-0.771 \pm 3.214$ &  $0.002 \pm 0.874$ \\
  \hline
\end{tabular}
\end{table}

\begin{table}[htp]
\caption{Estimated parameters $n_e$, $A^*$ and $K^*$ with the SDM
compared with the true parameters $n=0.26$, $\alpha$ and $K$ of the ETAS model
used to generate the synthetic catalogs using the event-by-event algorithm.
The other parameters are fixed to $m_d=0$, $b=1$,
$c=0.002$, and $\theta=0.19$.}
\label{tab_NewMAlpha}
  \begin{tabular}{ccrccccc}
  \# &\vline&  $n$  &   $n_{e}$   & $\alpha$ & $A^*$  &  $K$  & $K^*$ \\
  \hline
  1  &\vline&  0.26 & $0.244 \pm 0.024$ & 0.0 & $0.042 \pm 0.092$ & 0.260 & $0.845 \pm 0.211$ \\
  2  &\vline&  0.26 & $0.238 \pm 0.027$ & 0.1 & $0.033 \pm 0.082$ & 0.234 & $0.856 \pm 0.203$ \\
  3  &\vline&  0.26 & $0.248 \pm 0.022$ & 0.2 & $0.016 \pm 0.077$ & 0.208 & $0.895 \pm 0.176$ \\
  4  &\vline&  0.26 & $0.246 \pm 0.036$ & 0.3 & $-0.014 \pm 0.075$ & 0.182 & $0.940 \pm 0.173$ \\
  5  &\vline&  0.26 & $0.233 \pm 0.011$ & 0.4 & $-0.032 \pm 0.070$ & 0.156 & $0.967 \pm 0.137$ \\
  6  &\vline&  0.26 & $0.234 \pm 0.024$ & 0.5 & $-0.025 \pm 0.072$ & 0.130 & $0.960 \pm 0.152$ \\
  7  &\vline&  0.26 & $0.233 \pm 0.017$ & 0.6 & $-0.101 \pm 0.065$ & 0.052 & $1.069 \pm 0.107$ \\
  8  &\vline&  0.26 & $0.216 \pm 0.024$ & 0.7 & $-0.074 \pm 0.060$ & 0.104 & $1.011 \pm 0.116$ \\
  9  &\vline&  0.26 & $0.206 \pm 0.021$ & 0.8 & $0.179 \pm 0.952$ & 0.078 & $1.005 \pm 0.314$ \\
  10 &\vline&  0.26 & $0.216 \pm 0.210$ & 0.9 & $-0.121 \pm 0.055$ & 0.026 & $1.151 \pm 0.205$ \\
  \hline
\end{tabular}
\end{table}

\begin{table}[htp]
\caption{Estimated parameters $n_e$, $A^*$ and $K^*$ with the SDM
compared with the true parameters $n=0.6$, $\alpha=0.2$ and $K=0.48$ of the ETAS model
used to generate the synthetic catalogs by the event-by-event algorithm,
for various magnitude threshold $m_d$
of incompleteness. The other parameters have been fixed to $b=1$,
$c=0.001$, and $\theta=0.1$.}
\label{tab_NewMMd}
  \begin{tabular}{cccrccc}
  \# &\vline&  $m_d$  & $N_{m_d}$ & $n_{e}$    & $A^*$    & $K^*$ \\
  \hline
  1  &\vline&  0.0 & 4000 & $0.445 \pm 0.012$  & $-0.004 \pm 0.048$ &  $0.960 \pm 0.115$ \\
  2  &\vline&  0.5 & 1300 & $0.208 \pm 0.012$  & $0.000 \pm 0.0345$ &  $0.968 \pm 0.098$ \\
  3  &\vline&  1.0 &  400 & $0.080 \pm 0.014$  & $-0.004 \pm 0.022$ &  $1.013 \pm 0.085$ \\
  4  &\vline&  1.5 &  130 & $0.024 \pm 0.012$  & $0.003 \pm 0.017$ &  $0.977 \pm 0.073$ \\
  5  &\vline&  2.0 &   40 & $0.003 \pm 0.009$  & $0.009 \pm 0.018$ &  $0.932 \pm 0.077$ \\
  \hline
\end{tabular}
\end{table}

\begin{table}[htp]
\caption{Estimated parameters $n_e$, $A^*$ and $K^*$ with the MISD method
compared with the true parameters $n=0.4$, $\alpha=0.2$ and $K=0.32$ of the ETAS model
used to generate the synthetic catalogs by the generation-by-generation algorithm,
for various magnitude threshold $m_d$
of incompleteness. The other parameters have been fixed to $b=1$,
$c=0.001$, and $\theta=0.1$.}
\label{tab_misd21}
  \begin{tabular}{cccrccc}
  \# &\vline&  $m_d$  & $N_{m_d}$ & $n_{e}$    & $A^*$    & $K^*$ \\
  \hline
  1  &\vline&  0.0 & 2000 & $0.365 \pm 0.082$  & $0.380 \pm 0.198$ &  $0.618 \pm 0.393$ \\
  2  &\vline&  0.5 & 2000 & $0.235 \pm 0.079$  & $0.275 \pm 1.462$ &  $0.324 \pm 0.450$ \\
  3  &\vline&  1.0 & 1000 & $0.170 \pm 0.077$  & $0.041 \pm 2.094$ &  $0.348 \pm 0.597$ \\
  4  &\vline&  1.5 &  500 & $0.176 \pm 0.074$  & $0.065 \pm 1.501$ &  $0.364 \pm 0.651$ \\
  5  &\vline&  2.0 &  150 & $0.193 \pm 0.126$  & $0.114 \pm 0.991$ &  $0.411 \pm 0.621$ \\
  \hline
\end{tabular}
\end{table}

\begin{table}[htp]
\caption{Estimated parameters $n_e$, $A^*$ and $K^*$ with the SDM and MISD
compared with the true parameters $n$, $\alpha$ and $K$ of the ETAS model
used to generate the synthetic catalogs by the event-by-event algorithm. The other parameters have been
fixed to $m_d=m_0=0$, $b=1$, $c=0.001$, and $\theta=0.5$.}
\label{tab_add1}
  \begin{tabular}{rcccccccc}
  \# & method &\vline& $n$ & $\alpha$ & $K$ &    $n_{e}$      &     $A^*$     &    $K^*$ \\
  \hline
  1 & SDM   &\vline&  0.2 & 0.2  & 0.16  & $0.199 \pm 0.009$ & $-0.020 \pm 0.042$ & $0.999 \pm 0.087$ \\
  2 & MISD  &\vline&  0.2 & 0.2  & 0.16  & $0.313 \pm 0.019$ & $1.116 \pm 1.203$  & $0.367 \pm 0.285$ \\
  \hline
  3 & SDM   &\vline&  0.5 & 0.5  & 0.25  & $0.502 \pm 0.020$ & $-0.061 \pm 0.076$ & $1.004 \pm 0.156$ \\
  4 & MISD  &\vline&  0.5 & 0.5  & 0.25  & $0.546 \pm 0.018$ & $0.970 \pm 0.828$  & $0.501 \pm 0.363$ \\
  \hline
  5 & SDM   &\vline&  0.8 & 0.8  & 0.16  & $0.698 \pm 0.072$ & $0.024 \pm 0.481$  & $0.941 \pm 0.250$ \\
  6 & MISD  &\vline&  0.8 & 0.8  & 0.16  & $0.755 \pm 0.045$ & $0.700 \pm 0.108$  & $0.347 \pm 0.209$ \\
  \hline
  7 & SDM   &\vline&  0.8 & 0.2  & 0.64  & $0.793 \pm 0.014$ & $0.009 \pm 0.068$  & $0.942 \pm 0.141$ \\
  8 & MISD  &\vline&  0.8 & 0.2  & 0.64  & $0.804 \pm 0.016$ & $0.352 \pm 0.337$  & $0.626 \pm 0.272$ \\
  \hline
  9 & SDM   &\vline&  0.2 & 0.8  & 0.04  & $0.168 \pm 0.021$ & $-0.160 \pm 0.081$ & $1.187 \pm 0.159$ \\
  10 & MISD &\vline&  0.2 & 0.8  & 0.04  & $0.259 \pm 0.037$ & $0.694 \pm 0.128$  & $0.175 \pm 0.132$ \\
  \hline
\end{tabular}
\end{table}

which 

\end{document}